\documentclass[11pt,graphics]{article}

\usepackage{epsfig}
\usepackage{graphicx}
\usepackage{amssymb}
\usepackage{here}

\evensidemargin=\oddsidemargin

\newcommand{\beq}{\begin{equation}}
\newcommand{\eeq}{\end{equation}}
\newcommand{\bq}{\begin{equation}}
\newcommand{\eq}{\end{equation}}
\newcommand{\ba}{\begin{array}}
\newcommand{\ea}{\end{array}}
\newcommand{\beqa}{\begin{eqnarray}}
\newcommand{\eeqa}{\end{eqnarray}}

\newcommand{\tx}{\texttt}

\def\End{\end{document}}

\def\del{\partial }
\def\to{\rightarrow}

\def\dis{\displaystyle}
\def\f{\frac}

\def\[{\left[}
\def\]{\right]}
\def\({\left(}
\def\){\right)}

\def\U1EM{U(1)_{\rm em}}

\def\STU{{(S,T,U)}}
\def\T{{\cal T}}
\def\leqq{\leqslant}

\def\K{\kappa}
\def\DK{\Delta\kappa}

\def\[{\left[}
\def\]{\right]}
\def\dis{\displaystyle}

\def\cut{\Lambda}

\def\N{{\cal N}}

\def\thisday{November, 2002}

\def\ifb{ {\rm fb}^{-1} }

\begin{document}
\thispagestyle{empty}

{{\large \thisday} \hfill {\large\hfill {hep-ph/0211229}}
\vspace*{15mm}

\begin{center}
{\Large {\bf
Anomalous Gauge Interactions of the Higgs Boson: \\[2mm]
Precision Constraints and Weak Boson Scatterings
}}

\vspace*{15mm}
{\large\sc Hong-Jian He}\,$^{\rm a,}$\footnote{
Electronic address: hjhe@physics.utexas.edu},~~\,
{\large\sc Yu-Ping Kuang}\,$^{\rm b,c,}$\footnote{
Electronic address: ypkuang@mail.tsinghua.edu.cn},~~\,
{\large\sc C.--P. Yuan}\,$^{\rm d,}$\footnote{
Electronic address: yuan@pa.msu.edu},~~\,
{\large\sc Bin Zhang}\,$^{\rm c,}$\footnote{
Electronic address: zhang$_{-}\!$bin@mails.tsinghua.edu.cn}

\vspace*{4mm}
$^{\rm a}$\,{\rm
      Center for Particle Physics and Department of Physics,\\
      University of Texas at Austin, Texas 78712, USA
}
\\[3mm]
$^{\rm b}$\,{\rm
China Center of Advanced Science and Technology (World Laboratory),\\
P. O. Box 8730, Beijing 100080, China
}
\\[3mm]
$^{\rm c}$\,{\rm
Department of Physics, Tsinghua University, Beijing 100084, China
}
\\[3mm]
$^{\rm d}$\,{\rm
Department of Physics and Astronomy,\\
Michigan State University, East Lansing, Michigan 48824, USA
}
\end{center}

\vspace*{20mm}
\begin{abstract}
\noindent
Interaction of Higgs scalar ($H$) with weak gauge bosons
$(V=W^\pm,\,Z^0)$ is the {\it key} to understand
electroweak symmetry breaking (EWSB) mechanism.
New physics effects in the $HVV$ interactions, as predicted by
models of compositeness, supersymmetry and extra dimensions,
can be formulated
as anomalous couplings via a generic effective Lagrangian.
We first show that the existing electroweak precision data already
impose nontrivial indirect constraints on the anomalous $HVV$
couplings. Then, we systematically study
$VV \to VV$ scatterings in the TeV region,
via Gold-plated pure leptonic decay modes of the weak bosons.
We demonstrate that,
even for a light Higgs boson in the mass range
\,$115\,{\rm GeV} \lesssim m_H^{~} \lesssim 300$\,GeV,\,
this process can directly probe the anomalous
$HVV$ interactions at the LHC
with an integrated luminosity of 300\,fb$^{-1}$,  which
further supports the ``no-lose'' theorem for the LHC to
uncover the EWSB mechanism.
Comparisons with the constraints from measuring
the cross section of $VH$ associate production and
the Higgs boson decay width are also presented.
\\[4mm]
{PACS number(s): 14.65.Ha, 12.15.Lk, 12.60.Nz}
\\[2mm]
Preprint number(s): UT-HEP-02-08,  TUHEP-TH-01131, MSUHEP-20425
\end{abstract}

\newpage
\setcounter{page}{1}
\setcounter{footnote}{0}
\renewcommand{\thefootnote}{\arabic{footnote}}

\def\doublespaced{\baselineskip=\normalbaselineskip\multiply\baselineskip
  by 120\divide\baselineskip by 100}

\section{\hspace*{-6mm}.$\!$  Introduction}

Unraveling the electroweak symmetry breaking (EWSB) mechanism is
the most pressing task for the experiments at the TeV energy
colliders, such as the Fermilab Tevatron, the CERN Large Hadron
Collider (LHC), and the future Linear Colliders (LC). The Standard
Model (SM), with a single Higgs boson ($H$), provides the simplest
realization of the EWSB, which however is plagued with many
diseases (such as the triviality  
and the hierarchy problem)  
that have intrigued a number of attractive
resolutions including new physics models with dynamical symmetry
breaking\,\cite{DSB},
weak scale supersymmetry\,\cite{susy}, and
large or small extra dimensions\,\cite{extraD}.
If a relatively light Higgs boson is found at these colliders, its
gauge interaction with weak gauge bosons ($V=W^\pm,\,Z^0$) should be
quantitatively tested as the {\it key} to uncover the mechanism of
the EWSB. For instance, deviations in the $HVV$ couplings
naturally arise from the composite Higgs models\,\cite{GK84,DSB},
the supersymmetry models\,\cite{susy}, the extra dimension models
(with Higgs on or off the standard model brane)\,\cite{RW},
and the deconstruction models (with ``little Higgs'' from the
theory space)\,\cite{DeC}.
In this Letter, we first analyze how the updated electroweak precision data
already impose nontrivial constraints on the anomalous
$HVV$ couplings.  Then, we propose to test
the anomalous $HVV$ couplings at the LHC by quantitatively studying the
weak gauge boson scatterings in the TeV regime.
We demonstrate that even for a light Higgs boson
in the mass range
\,$115\,{\rm GeV} \lesssim m_H^{~} \lesssim 300$\,GeV,\,
this process can sensitively test the $HVV$
interactions, which further supports
the ``no-lose'' theorem\,\cite{CG}
for the LHC to probe the EWSB mechanism.
Finally, for comparison, we examine the
sensitivity of high energy colliders to the $HVV$ coupling
from the associate $VH$ production,
as well as the total decay width of the Higgs boson.

The Standard Model (SM) is an effective theory,
valid only up to certain energy scale $\cut$, below which
all new physics effects can be parametrized as appropriate
effective operators in terms of the SM fields. 
The Higgs sector can be economically formulated by the nonlinear
realization\,\cite{CCWZ,Wein,App,CK}, which is particularly
convenient when the EWSB invokes strong dynamics\,\cite{DSB}.
Such an effective Lagrangian was explicitly constructed in 
Ref.\,\cite{CK}, containing
the nonlinearly realized Higgs boson field $H$
(transforming as a weak singlet with its mass $m^{~}_H<\Lambda$), the
triplet would-be Goldstone boson fields ${\overrightarrow \omega}$, and
the electroweak gauge boson fields.
Under the electroweak gauge symmetry, assuming
the invariance of the charge conjugation {\tt C} and the parity
{\tt P}, as well as the custodial $SU(2)_c$ symmetry
(violated only by $g'\neq 0$),
the effective Lagrangian can be written as,
up to dimension-4,\footnote{An extension
of our study to include the linearly realized
effective Higgs Lagrangian\,\cite{BW} will be presented
elsewhere\,\cite{ZKHY}.}
\begin{eqnarray}                  
{\cal L}_{\rm eff}^{(d\leqslant 4)}
&=&-\frac{1}{4}{\overrightarrow
W}_{\mu\nu}\cdot{\overrightarrow
W}^{\mu\nu}-\frac{1}{4}B_{\mu\nu}B^{\mu\nu}  
+\frac{1}{4}\(v^2+2\K vH+\K' H^2   
\){\rm Tr}(D_\mu\Sigma^\dagger D^\mu\Sigma)
\nonumber\\
&&+\frac{1}{2}(\del_\mu H)(\del^\mu H)
-\frac{m_H^2}{2}H^2-\frac{\lambda_3 v}{3!}H^3+\frac{\lambda_4}{4!}H^4,
\label{Lagrangian}
\end{eqnarray}
where $\overrightarrow W_{\mu\nu}$ and $B_{\mu\nu}$ are field
strengths of the $SU(2)_L$ and $U(1)_Y$
electroweak gauge fields, respectively; $v\simeq 246$\,GeV is the
vacuum expectation value characterizing the EWSB;
$(\K,\,\K',\,\,\lambda_3,\,\lambda_4)$ are the anomalous
coupling constants, and
\begin{eqnarray}                    
\Sigma=\displaystyle
\exp [i\overrightarrow\tau\cdot{\overrightarrow\omega}/v]\,,
~~~~~
D_\mu\Sigma=\partial_\mu\Sigma+i\f{g}{2}{\overrightarrow\tau}\cdot
{\overrightarrow
W}_\mu\Sigma -i\f{g'}{2}B_\mu\Sigma{\tau_3}\,,\nonumber
\end{eqnarray}
in which the Pauli matrix $\tau_i$ is normalized as
${\rm Tr}(\tau_i,\tau_j)=2\delta_{ij}$.
In terms of the above notations,
the SM Higgs boson has the interactions
corresponding to $\K=\K'=1$ and
$\lambda_3=\lambda_4 = 3 m_H^2/v^2$.
In the current study, we assume that except $H$,
all the other Higgs scalars, if exist,
are heavy and around the scale $\cut$ or above,
so that only $H$ is relevant to the effective theory.

\section{\hspace*{-6mm}.$\!$
Constraints from Precision Electroweak Data}

As indicated in Eq.\,(\ref{Lagrangian}), there is an
important difference between the nonlinearly and the linearly
realized Higgs sector.
The nonlinear formalism allows new physics to appear in
the effective operators with dimension $\leqq 4$ whose
coefficients are not necessarily
suppressed by the cutoff scale $\Lambda$.
Hence, the couplings of Higgs boson to weak gauge bosons can
naturally deviate from the SM-values by
an amount of $\lesssim {\cal O}(1)$,
according to the naive dimensional analysis \cite{NDA}.

In the unitary gauge, the relevant anomalous $HVV$ couplings to
the precision oblique parameters $(S,\,T,\,U)$\,\cite{PT}
take the following form:
\beq                               
\dis
\[(\K-1)2vH +(\K'-1)H^2\]
\[\f{2m_W^2}{v^2}W^+ W^-
 +\f{m_Z^2}{v^2} Z Z \] \, .
\eeq
The deviations of $\K$ and $\kappa^{\prime}$ from 1 represent the
effect from new physics.
Hereafter, we define
\,$\Delta \K\equiv\K-1$\, and \,$\Delta \K'\equiv\K'-1$.\,
When calculating radiative corrections using the
effective Lagrangian (\ref{Lagrangian}), it is generally
necessary to introduce higher dimensional counter terms
to absorb new divergences arising from the loop integration. There are
in principle three next-to-leading order (NLO) counter terms to render
the $\STU$ parameters finite at the one-loop level, called ${\cal
L}^{(2)^\prime},\,{\cal L}^{(4)}_1,\,{\cal L}^{(4)}_8$ \cite{App},
i.e.,
\beq                         
\label{eq:L018}
\dis \ell_0 \f{v^2}{16\pi^2}
\f{1}{4}
\[{\rm Tr}{\T} (D_\mu{\Sigma}){{\Sigma}}^\dagger\]^2,~~~
\ell_1 \f{v^2}{\cut^2} gg'
{\rm Tr}\[{\bf B}_{\mu\nu}\Sigma^\dagger {\bf W}^{\mu\nu}\Sigma\],~~~
\dis \ell_8 \f{v^2}{\cut^2}\f{g^2}{4}
\[{\rm Tr}(\T {\bf W}_{\mu\nu})\]^2,
\eeq
whose coefficients $(\ell_0,\ell_1,\ell_8)$ correspond to
the oblique parameters $(T,\,S,\,U)$, respectively\footnote{
Comparing to ${\cal L}^{(4)}_1$, ${\cal L}^{(4)}_8$
has the same dimension, but contains two new
$SU(2)_c$-violating operators of $\T$, so that we expect
$\ell_8/\ell_1 \sim 1/16\pi^2 \sim 10^{-2} \ll 1$. This generally
leads to \,$U\ll (S,\,T)$\,.}.
Here ${\bf W}_{\mu\nu} ={\overrightarrow W}_{\mu\nu}\!\cdot\!
{\overrightarrow  \tau}/2$, ${\bf B}_{\mu\nu}
={B}_{\mu\nu}\tau_3/2$, and ${\cal
T}\equiv\Sigma\tau_3\Sigma^\dagger$. To estimate the contribution of
loop corrections, we invoke a naturalness assumption 
that no fine-tuned accidental cancellation occurs between the
leading logarithmic term and the constant piece of the counter
terms. Thus, the leading logarithmic term represents a reasonable
estimate of the loop corrections\footnote{ This approach is commonly
used in the literature for estimating new physics effects in
effective theories\,\cite{eg}.}.   It is straightforward to compute
the radiative corrections to $\STU$ due to the $HVV$ anomalous
couplings, under the $\overline{\rm MS}$ scheme, using dimensional
regularization and keeping only the leading logarithmic terms.
After subtracting the SM Higgs contributions ($\K=1$) at the
reference point $m_H^{\rm ref}$, we find,\footnote{At the
one-loop order, the coupling $\K'$ has no contribution to
the $S$, $T$ and $U$ parameters.}
\beq                        
\label{eq:STUH}
\ba{l}
\hspace*{-5mm}
\Delta S = \dis
\f{1}{6\pi}\[\ln\f{m^{~}_H}{\,m_H^{\rm ref}}
     -(\K^2\!-\!1)\ln\f{\Lambda}{\,m^{~}_H} \]\!,~~
\Delta T = \dis
\f{3}{8\pi c_{\rm w}^2}
\[-   \ln\f{m^{~}_H}{\,m_H^{\rm ref}}
+(\K^2\!-\!1)\ln\f{\Lambda}{\,m^{~}_H}\]\!,~~ \Delta U = \dis 0\,,
\ea \eeq where the $\ln\cut$ term represents the genuine new
physics effect arising from physics above the cut-off scale
$\cut$\,\cite{BL}. For a SM Higgs boson ($\K=1$), a heavier Higgs
mass will increase $\Delta S$ and decrease $\Delta T$. Choosing
the reference Higgs mass $m_H^{\rm ref}$ to be $m^{~}_H$ can
further simplify Eq.\,(\ref{eq:STUH}) as
\beq                       
\label{eq:STUf}
\ba{l}
\Delta S = \dis
  -\f{\,\K^2-1\,}{6\pi}\ln\f{\Lambda}{\,m^{~}_H}\,,~~~
\Delta T = \dis
  +\f{\,3(\K^2-1)\,}{8\pi c_{\rm w}^2}\ln\f{\Lambda}{\,m^{~}_H}\,,~~~
\Delta U = \dis 0 \, .
\ea
\eeq
For a given value of $\cut$ and $m^{~}_H$,
when $|\K| >1$, we have
  $\Delta S < 0$ and $\Delta T> 0$.
This pattern of radiative corrections
allows a relatively heavy Higgs
boson to be consistent with the current precision data
(cf. Fig.\,1a).
Moreover, Eq.\,(\ref{eq:STUf}) also indicates that the loop
contribution from $\K\neq 1$ induces a sizable
ratio of \,$\Delta T/\Delta S = -9/(4c_w^2) \approx -3$\,.

\begin{figure}[H]
\vspace*{-10mm}
\hspace*{1cm}
\centerline{\epsfig{figure=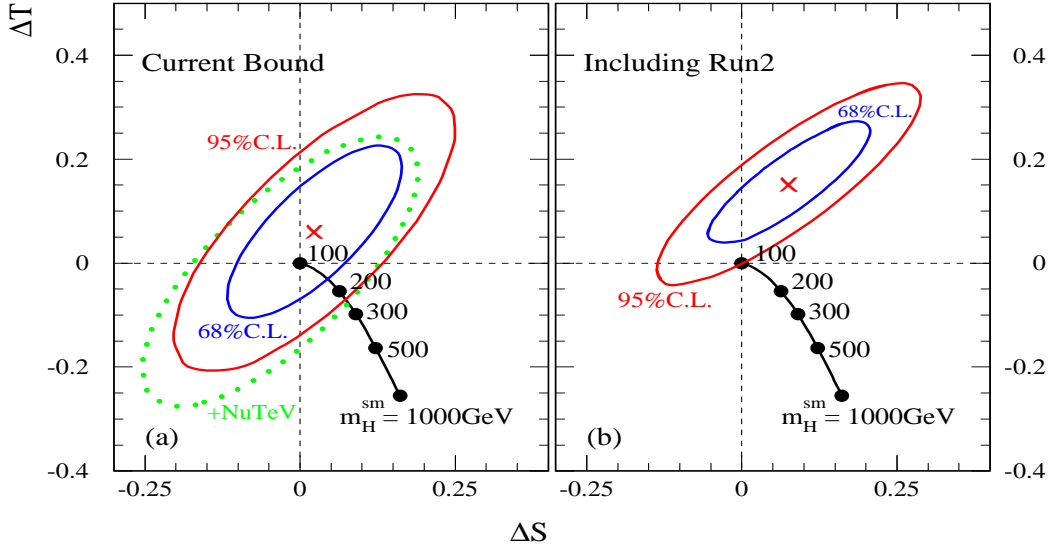,height=9cm,width=16.5cm}}
\vspace{-1cm}
\caption{$\Delta S -\Delta T$ contours (a) from the current
precision electroweak data, and (b)
from including the expected Tevatron Run-2
sensitivity to $m^{~}_W$ and $m^{~}_t$
[assuming the current central values of $(m^{~}_W,\,m^{~}_t)$ with
an error of $(20\,{\rm MeV},\,2$\,GeV)].
Here, we set $m_H^{\rm ref}=100$\,GeV and $\Delta U=0$ in the
global fit.
The dotted curve in (a) is the 95\%\,C.L. bound after including the
NuTeV data.
}
\label{Fig:Fig1}
\end{figure}

If one {\it assumes} that there were no new physics beyond the
SM,  a global fit to the current precision electroweak data would
suggest the SM Higgs to be light, with a central value
$m^{~}_H=83$\,GeV (significantly below the LEP2 direct search
limit $m^{~}_H > 114.3$\,GeV \cite{PDG}) and a 95\%\,C.L. limit
on the range of Higgs boson mass: $\,32\,{\rm GeV}\,\leqq m^{~}_H
\leqq 192$\,GeV.\,  However, it was recently pointed out that in
the presence of {\it new physics,}  such a bound can be
substantially relaxed \cite{seesaw,bagger,HPS,snow,Mike}. In the
above fit we did not include the latest NuTeV data. If the NuTeV
data is included, the value of the minimum $\chi^2$ of the global
fit increases substantially (by $8.7$), indicating a poor quality
of the SM fit to the precision data. (This fit gives a similar
central value,  \,$m^{~}_H=85$\,GeV,\, and 95\%\,C.L. mass range
$\,33\,{\rm GeV}\,\leqq m^{~}_H \leqq 200$\,GeV.)  A similar
increase of $\chi^2$ (by 8.9) also appears in the $\Delta S-\Delta T$
fits, suggesting that the NuTeV anomaly cannot be explained by the
new physics effect arising from the oblique parameters $(\Delta
S,\,\Delta T)$ alone. Since the potential problems with the NuTeV
analysis are still under debate\,\cite{NuTeVx}, the NuTeV data will
not be included in the following analysis\footnote{
For comparison, in Fig.\,1(a) we have displayed a 95\%\,C.L. contour
(dotted curve) from the fit by including the NuTeV anomaly.}. 
In Fig.\,1(a), we show the
$\Delta S-\Delta T$ bounds (setting $m_H^{\rm ref} = 100$\,GeV),
derived from the global fits with
the newest updated electroweak precision data\,\cite{elgapp,data-new}.
Furthermore, for $m_H^{\rm ref} = 115\,(300)$\,GeV and
$\Delta U=0$, the global fits give 
\begin{eqnarray}            
\Delta S=0.01\,(-0.07)\pm 0.09\,,
~~~~~
\Delta T=0.07\,(0.16)\pm 0.11 \,.
\label{ST}
\end{eqnarray}
What also shown in the same figure is the contribution to $\Delta
S$ and $\Delta T$ from the SM Higgs boson with different masses.
Fig.\,1(b) shows that the upcoming measurements of
the $W^\pm$ mass ($m_W$) and top mass ($m_t$)
at the Tevatron Run-2 can significantly improve the constraints
on the new physics via the oblique corrections,
where the current Run-1 central values of $(m^{~}_W,\,m^{~}_t)$
are assumed, but with their errors
reduced to the planned sensitivity of $20$\,MeV and $2$\,GeV, respectively.

\begin{figure}[H]
\vspace*{-13mm}
\begin{center}
\hspace*{15mm}
\centerline{\epsfig{figure=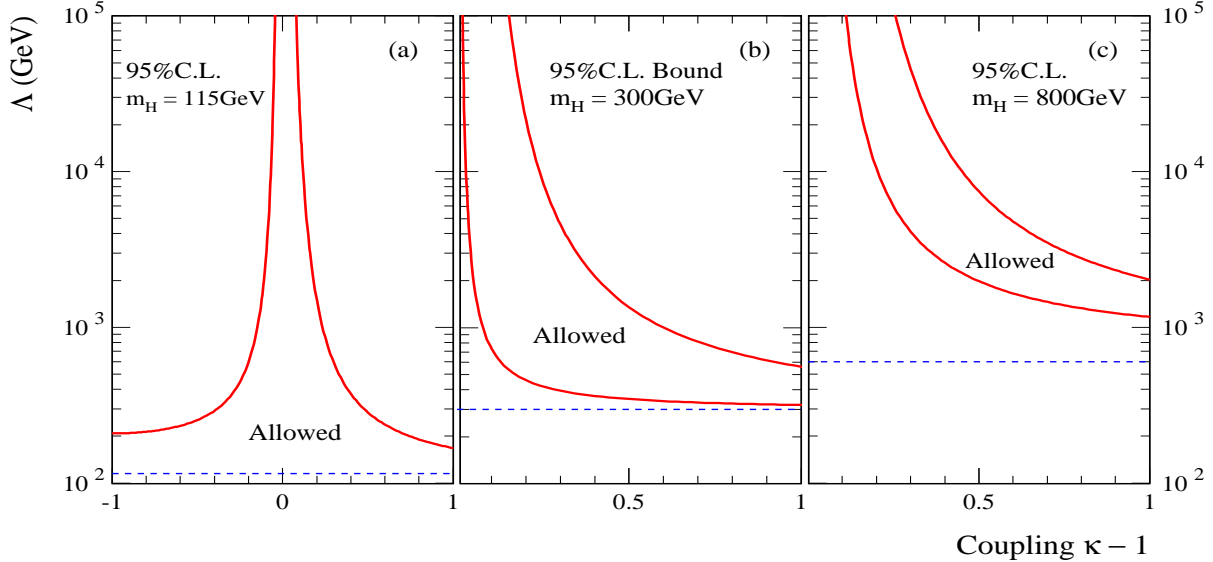,height=9cm,width=20cm}}
\vspace*{-13mm}
\end{center}
\caption{Constraints on the new physics scale $\Lambda$
as a function of the anomalous coupling $\Delta \kappa$.
The regions below the solid curves and above the dashed lines
in (a) or between the two solid curves in (b)-(c)
are allowed at the $95\%$C.L. The dashed lines indicate the value
of Higgs mass $m^{~}_H$.
}
\label{Fig:Fig2}
\end{figure}

Given the allowed range of $(\Delta S,\,\Delta T)$, as shown in
Fig.\,1(a), we can further constrain the new physics scale
$\Lambda$ as a function of the anomalous coupling $\Delta\K $
for a given $m_H^{\rm ref}=m^{~}_H$ value [cf.
Eq.(\ref{eq:STUf})]. The results are depicted in Fig.\,2.
Alternatively, from Fig.\,2, we can constrain the range of
$\Delta\kappa$ for given values of $(\cut,\,m^{~}_H)$, as
summarized in Table\,1.
\vspace*{-3mm}
\begin{table}[H]
\label{Tab:Tab1}
\caption{The 95\%\,C.L. limits on $\DK$ for typical values of
new physics scale  $\cut$ and Higgs mass $m_H^{~}$.
}
\vspace*{-1mm}
\begin{center}
\begin{tabular}{r|ccc}
\hline\hline
&&&\\[-2.5mm]
$\cut=$~\,  &  $1$\,TeV  & $10$\,TeV   &  $100$\,TeV
\\ [1.5mm]
\hline 
&&&\\[-2.5mm]
~~$m_H=115$\,GeV~\,
& ~$-0.15\leqq \DK \leqq 0.23$~
& ~$-0.069\leqq \DK \leqq 0.12$~
& ~$-0.045\leqq \DK \leqq 0.08$~\,
\\ 
$300$\,GeV~\,
& $-0.074\leqq \DK \leqq 0.60$
& $ 0.027\leqq \DK \leqq 0.24$
& $ 0.016\leqq \DK \leqq 0.15$
\\ 
$800$\,GeV~\,
& (excluded)
& $ 0.20\leqq \DK \leqq 0.45$
& $ 0.11\leqq \DK \leqq 0.26$
\\[1.5mm]
\hline\hline
\end{tabular}
\end{center}
\end{table}
\vspace*{-4mm} Fig.\,2 and Table\,1 show that for
$\,m^{~}_H\gtrsim 250-300$\,GeV, the $\,\DK < 0\,$ region is
fully excluded, while a sizable $\,\DK > 0\,$ is allowed provided
$\cut$ is relatively low. Furthermore, for $\,m^{~}_H\gtrsim
800$\,GeV, the region $\,\cut<1.1$\,TeV is excluded. For
$\,m^{~}_H\gtrsim 250-300$\,GeV, the preferred range of $\DK > 0$
requires the couplings of $HW^+W^-$ and $HZZ$ to be stronger than
that of the SM, so that the direct production rate of the Higgs
boson, via either the Higgs-strahlung or the $VV$
fusions in high energy collisions, should raise above the SM rate.
On the other hand, for $m^{~}_H\lesssim 250$\,GeV, the direct
production rate of the Higgs boson can be smaller or larger than
the SM rate depending on the sign of $\Delta \K$.   Thus, when
$\Delta \K < 0$, a light non-standard Higgs boson may be
partially {\it hidden} by its large SM background events.
However, in this case, the new physics scale $\cut$ will be
generally low. For instance, when $m^{~}_H=115$\,GeV, a negative
$\DK = -0.15\,(-0.28)$ already forces $\cut\leqq 1\,(0.4)$\,TeV.
Finally, we note that for certain class of models the new physics
may also invoke extra heavy fermions such as in the typical
top-seesaw models with new vector-like fermions \cite{seesaw,DSB}
or models with new chiral families \cite{HPS}. In that case,
there can be generic positive contributions to $\Delta T$, so
that the $\DK < 0$ region may still be allowed for a relatively
heavy Higgs boson, but such possibilities are very
model-dependent. In our current effective theory analysis, we
consider the bosonic $HVV$ couplings as the dominant
contributions to the oblique parameters, and assume that other
possible anomalous couplings (such as the deviation in the gauge
interactions of $tbW/t\bar{t}Z$) may be ignored.
However, independent of these assumptions, the {\it most decisive
test} of the $HVV$ couplings
can come from direct measurements via
Born-level processes at the high energy colliders, which is the
subject of the next two sections.

\section{\hspace*{-6mm}.$\!$
Probing $HVV$ Interaction from
Weak Boson Scattering}

When a light Higgs boson (less than about $300$\,GeV)
is detected, the anomalous coupling $\kappa$ may be
measured from the production and the decay of the Higgs boson
(cf. next section). Moreover, it is important to study the
weak-boson scatterings in the TeV region and test whether this
Higgs boson is truly responsible for the mechanism of
the spontaneous EWSB,
i.e., for generating the longitudinally polarized
weak bosons (and their observed masses).
Consider the effective Lagrangian (\ref{Lagrangian}).
When $\kappa=1$, the weak gauge boson scatterings
in the TeV region will be dominated by the transversely polarized
weak bosons (denoted as $V_T$) and the scattering amplitudes will
be small when the Higgs boson is light.
However, when $\kappa \neq 1$ and
$|\Delta \kappa | ={\cal O}(1)$,
the weak boson scatterings in the TeV regime
will be dominated by the longitudinally polarized weak bosons
(denoted as $V_L$)
and the scattering amplitudes will grow as the invariant
mass of the weak boson pair (denoted as $E$) increases.
Applying the electroweak power counting method\,\cite{power},  
we find that
the $V_TV_T$ scattering amplitude behaves as \,$g^2 \K^2 E^0$\,
and the $V_LV_L$ scattering amplitude rapidly grows as
\,$(\K^2-1)E^2/v^2$.\, For a large $E$ and $\DK \neq 0$, the
$V_LV_L$ scattering can dominate the $VV\to VV$ processes,
and the growth of the event rate can sensitively probe the
deviation $\DK$.

As shown in  Fig.\,2, for $m^{~}_H\gtrsim 250-300$\,GeV, the
precision electroweak data already exclude a negative $\Delta \K$.
Hence, if new physics effect sets in the $HVV$ coupling
$\K=1+\DK$, the production rates of $VH$ and $qq \to qqH$, and the
total decay width of $H$ should raise above the SM predictions.
It can be the case that for a sizable $\DK >0$, the total width of
$H$ becomes so large that $H$ cannot be detected as a sharp
resonance\,\cite{nsH2} and therefore escapes the detection when
scanning the invariant mass of its decay products around $m^{~}_H$
region. In this case, can we still probe such a sub-TeV Higgs
boson? The answer is yes: as explained above, it unavoidably leads
to enhanced $VV$-scattering amplitudes of ${\cal
O}\!\((\K^2-1)E^2/v^2\)$ in the {\it TeV region}, detectable at
the LHC. On the other hand, if $\Delta\kappa$ is considerably
negative, the on-shell production rate of a light Higgs boson via
Higgs-strahlung or $VV$ fusion becomes so small that it may escape
the detections under these channels. However, in this case, the
$V_LV_L$ scattering amplitude has to still grow up with $E^2$ and become
large at the TeV scale due to the anomalous $\DK$ coupling!
Therefore, even if a light Higgs boson exists, studying the $VV$
scattering in the TeV regime remains important for testing the
true nature of its gauge interactions and the origin of the
spontaneous EWSB. Based upon the above observations, we propose to
directly test the anomalous coupling  $\Delta \K$ by studying $VV$
scatterings in the TeV region at the LHC. We will show that rather
sensitive tests of $\Delta \K$ can be performed by measuring the
cross sections of the longitudinal weak-boson scattering,
$V_LV_L\to V_LV_L$, especially $W^+_LW^+_L\to W^+_LW^+_L$.

The scattering amplitude of $VV \to VV$ contains two parts:
(i) the amplitude $T(V,\gamma)$ related solely to electroweak gauge
bosons, and (ii) the amplitude $T(H)$ related to the
Higgs boson exchanges.
In the SM ($\K=1$), the $HVV$ vertex contains the same gauge coupling
as in the non-Abelian interaction of the weak bosons.
At high energies, both
$T(V,\gamma)$ and $T(H)$ grow with $E^2$.
However, for $\K=1$, the $E^2$-dependent
pieces in the two amplitudes precisely
cancel in the sum $T(V,\gamma)+T(H)$
so that the total amplitude only has
$E^0$-dependence, respecting the unitarity of
the $S$-matrix.
For the non-standard Higgs boson,
$\K\neq 1$ originates from the new physics above $\Lambda$,
and the two $E^2$-dependent pieces do not cancel, making the total
amplitude grow as $\,(\K^2-1)E^2/v^2\,$ in the region below
$\Lambda$.
Such anomalous $E^2$-behavior (rather than $E^0$-behavior of the SM
amplitude) occurs only in the  $V_LV_L\to V_LV_L$ channel and
is rather sensitive to $\Delta \K$\,.

To analyze the LHC sensitivity to probing the anomalous $HVV$
coupling, we compute the cross sections for $pp\to jjVV$
numerically using the parton distribution functions CTEQ6L
\cite{cteq6}. We consider only the gold-plated (pure leptonic
decay) modes of the final state $V$'s in order to avoid the large
hadronic backgrounds at the LHC. Even in this case, there are
still several classes of large backgrounds to be eliminated,
including the {\it electroweak (EW) background}, the {\it QCD
background}, and the {\it top quark background} studied in
Refs.\,\cite{WW95,Yuan}.  Following Refs.\,\cite{WW95,Yuan}, we
require {\it forward-jet tagging}, {\it central-jet vetoing} and
{\it detecting isolated leptons, nearly back to back, with large
transverse momentum in the central rapidity region} to suppress
the backgrounds. Since we compute the exact tree-level amplitudes
of $pp\to jjVV$ without invoking the effective-$W$ approximation
(EWA), our numerical results are valid not only for large $|\DK|$
values but also for small $|\DK|$ region where the
$V_L$-contribution to $VV$-scatterings becomes comparable or
smaller than that of $V_T$ (for $m_H\lesssim 300$\,GeV) and the
signal detection is much harder.

Following the above procedures,
we compute the tree-level cross sections of the scattering processes
\,$pp(VV) \to VVjj \to \ell\ell\ell\ell jj$.\,
We find that the most sensitive channel
to determine $\Delta\K$ is
\,$W^+ _LW^+_L\to W^+_LW^+_L$\, due to its
small background rates.  In Table\,2, we summarize
the number of events (including both signals and backgrounds)
for this most sensitive channel \,$pp\to W^+ W^+ jj\to
\ell^+\nu \ell^+\nu jj$
in the range of \,$-0.4\leqq \Delta\kappa\leqq 0.4$\, and
\,$115\,{\rm GeV}\leqq m_H\leqq 300\,{\rm GeV}$,\,
with an integrated luminosity of 300\,fb$^{-1}$.

\vspace*{-3mm}
\begin{table}[h]
\tabcolsep 2.5pt
\centering{\caption{Number of events at the
LHC\,(300\,fb$^{-1}$) for \,$pp\to W^+ W^+ jj
\to \ell^+\nu\ell^+\nu jj$\,
($\ell = e,\mu$) with various values of $m^{~}_H$ (GeV) and
$\Delta\K$ in the non-standard Higgs model, 
where $\Delta\kappa=0$ corresponds to the SM. 
Values of the statistical significance
$\,\N_S/\sqrt{\N_B}$\, are shown in the parentheses.}
\vspace*{4mm}
{\small
\begin{tabular}{r|ccccccccccc}
\hline\hline
             & & & & & & & & & & & ~~~~~~\\[-3.2mm]
$\Delta\kappa=$
  &$-0.40$ & $-0.30$ & $-0.24$ & $-0.21$ &
   $-0.18$ & $0.00$  & $0.18$  & $ 0.21$ & $0.23$ &
   $0.30$  & $0.40$~  \\
\hline
& & & & & & & & & & & ~~~~~~\\[-3.2mm]
$m_H^{~} =$
 115~ & ~34(4.9) & 27(3.1) & 23(2.1) & 21(1.5) & 19(1.0) & ~~15~~
 & 23(2.1) & 25(2.6) & 27(3.1)
 & 37(5.7) & 58(11)~
\\
 130~ & ~34(4.9) & 27(3.1) & 23(2.1) & 21(1.5) & 19(1.0) & 15
 & 23(2.1) & 25(2.6) & 27(3.1)
 & 37(5.7) & 57(11)~
\\
 200~ & ~35(5.2) &28(3.4) & 24(2.3) & 23(2.1) & 21(1.5)& 15
 & 20(1.3) & 23(2.1) & 25(2.6)
 & 33(4.6) & 52(9.6)~
\\
 300~ & ~36(5.0) & 30(3.5) & 26(2.5) & 24(2.0) & 23(1.8) & 16 
 & 19(0.8) & 22(1.5) & 23(1.8)
 & 29(3.3) & 43(6.8)~
\\[1mm]
\hline\hline
\end{tabular}}}
\end{table}

From our analysis, we find that 
the kinematic cuts of Ref.\,\cite{WW95} can effectively suppress
the backgrounds relative to the signal
for the case with $\Delta\kappa$ significantly different from zero
since the $E^2$-dependence of the $W_LW_L$ amplitude enhances the
signal. However, for $\Delta\kappa$ close to zero, only the QCD and top
quark backgrounds are negligibly small, while the EW background
is still quite large compared to the signal. 
For instance, when $\DK =0$, we see that the ${15\sim16}$
SM-events for $m_H\leqq 300$ GeV in Table\,2 come essentially
from the $W_TW_L$ and $W_TW_T$ contributions (EW backgrounds).
So, under these cuts, the real background events are given by, 
\,${\N}_B = {\cal N} [\DK = 0] = {15\sim 16}$\,.\, 
We can then define the signal events for $\Delta\kappa\ne 0$ as 
\,${\N}_S = {\cal N} [\DK \ne 0]-{\N}_B$\,.\, 
To see the sensitivity of the LHC for discriminating the
cases between
$\DK \neq 0$ and $\DK =0$ (SM), we also show the statistical
significance \,$\N_S/\sqrt{\N_B}$\, in the parentheses in Table\,2.
The values of $\Delta\kappa$ corresponding to the
$2\sigma$ level of deviations from the SM are explicitly
displayed in Table\,2.
Hence, if the $\DK\neq 0$ effect is not detected, the LHC can
constrain the range of $\Delta\kappa$ to be about 
\begin{eqnarray}
\label{eq:LHC-B}
-0.2 ~<~ \Delta\kappa ~<~ 0.2  \,,
\end{eqnarray}
at the $2\sigma$ level.

Before concluding this section,
we discuss the possible unitarity violation in
the scattering process
\,$pp\to W^+W^+ jj\to \ell\nu \ell \nu jj$ (for $\K \neq 1$),
whose leading contribution comes
from the sub-process
$W^+_L W^+_L \to W^+_L W^+_L$ when the initial $W_L^+W_L^+$
are almost collinearly radiated from the incoming quarks
or antiquarks.
The scattering amplitude of $W^+_L W^+_L \to W^+_L W^+_L$
contributes to the isospin $I=2$ channel, and
in the high energy region ($E^2\gg M^2_W, m^2_H$),
is dominated by the leading $E^2$-contributions,
$\,T[I=2] \simeq (\K^2-1)E^2/v^2$.\,
According to the partial-wave analysis, its
$s$-wave amplitude \,$a^{~}_{I,J=2,0}$\, is given by
\beq
a^{~}_{20} ~\simeq~\dis \(\K^2-1\)\f{E^2}{\,16\pi v^2\,} \,,
\eeq
where $E=M_{VV}$.
The unitarity condition for this channel is,
$\,|\Re\mathfrak{e}\, a^{~}_{20}| < {2!}/{2}=1$\,,\,
and the factor $2!$ is due to the identical $W^+W^+$
in the final state. This results in a requirement,
\beq
\label{eq:UB-K}
\dis
\sqrt{1-\f{\,16\pi v^2\,}{E^2}} ~<~ |\K| ~<~
\sqrt{1+\f{\,16\pi v^2\,}{E^2}} \,.
\eeq
For instance, it constrains
\,$|\K| < 3.6$\, for \,$E=500$\,GeV,\, and
\,$0.5\,(0.8) < |\K| < 1.3\,(1.2)$\, for \,$E=2\,(3)$\,TeV.\,
Hence,  the expected sensitivity of the LHC to determining
$\DK$ [cf. Eq.\,(\ref{eq:LHC-B})]
is consistent with the unitarity limit
since the typical invariant mass of
the $W^+W^+$ pair, after the kinematic cuts, falls into the range
\,$500\,{\rm GeV} \leqq E \leqq 2\sim 3$\,TeV.
The contributions from higher invariant mass values are
severely suppressed by the parton luminosities\,\cite{WW95,power}
and thus negligible.

\section{\hspace*{-6mm}.$\!$
Other Measurements}

The anomalous $HVV$ coupling can also be measured from the
associate production of the Higgs boson with the weak boson,
as well as the total decay width of the Higgs boson.

\subsection{\hspace*{-5mm}.$\!$
Constraints from the $VH$ Associate Production}

The anomalous $HVV$ coupling can be directly measured
via the associate $VH$ production in the lepton
or hadron collisions, $\,e^-e^+,\,q{q'}\to VH$.
The LEP-2 Higgs search puts a direct bound,
$m_H^{\rm SM} > 114.3$\,GeV\,\cite{PDG}.
For a non-SM Higgs boson, this lower bound can
be weakened. For instance, in the supersymmetric SM, the
$HZZ$ coupling is smaller than its SM value by a factor
$\sin(\alpha-\beta)$ or $\cos(\alpha-\beta)$, depending on whether
$H$ is the lighter or heavier \tx{CP}-even state, and this lower
bound is reduced to about 90\,GeV\,\cite{PDG}.
If the SM Higgs boson weighs about 110\,GeV,
Tevatron Run-2 will be able to detect it.
Assuming an integrated luminosity of $10\,\ifb$,
the number of expected signal events is about 27 and the
background events about 258, according to the Table\,3
(the most optimal scenario) of Ref.\,\cite{agrawal}. Therefore, the
$1\sigma$ statistic fluctuation of the background event is
about $16$.
Consequently, we find, at the $1\sigma\,(2\sigma)$ level,
\,$0.6 < |\K|< 1.2$ ($|\K| < 1.5$)\,.\footnote{This
bound can be improved by carefully examining
the invariant mass distribution of the $b {\bar b}$ pairs
in the Higgs decay.}\,
Similarly, we can estimate the bounds on $|\K|$
for various $m^{~}_H$ values as below
\vspace*{-3mm}
\beq                      
\ba{rcc}
& 1\sigma~\, & 2\sigma~~~\\[1.5mm]
m_H^{~}=110~{\rm GeV:} &
~~~0.6 \leqq |\K|\leqq 1.2\,,~~~ &  0 \leqq |\K|\leqq 1.5\,;
\\[1mm]
m_H^{~}=120~{\rm GeV:}  &
0.4 \leqq |\K|\leqq 1.4\,, & 0 \leqq |\K| \leqq 1.6\,;
\\[1mm]
m_H^{~}=130~{\rm GeV:}   &
 ~~0 \leqq |\K|\leqq 1.5\,, & 0 \leqq |\K|\leqq 1.8\,.
\\
\ea
\label{tevbound}
\eeq
It is clear that the above limits can be further
improved at the Tevatron Run-2 by having a larger
integrated luminosity until the systematical
error dominates over the statistical error.
The same process can also be studied at the LHC to test
the anomalous coupling $\K$.
However, because of much larger background rate at the LHC, the
improvement on the measuring $\K$ via the $VH$ associate production
is not expected to be significant.

\subsection{\hspace*{-5mm}.$\!$
Constraints from Decay Width of Higgs Boson}

Another method to determine $\K$ is to measure the decay width of
the Higgs boson.  When \,$m^{~}_H > 2m^{~}_Z$,\,
the decay process $\,H\to ZZ\to 4\mu$\,
is one of the ``gold-plated'' channels
(the pure leptonic decay modes)
that allow the reconstruction of the $ZZ$
invariant mass with a high precision.
Thus, the total decay width of the Higgs boson
can be precisely measured.
A detailed Monte Carlo analysis for such a measurement
at the LHC was carried out in Ref.\,\cite{width}.
Assuming
that there is no non-SM decay channel open except the presence
of the anomalous $HVV$ coupling $\K$, we can directly convert the
results of Ref.\,\cite{width}  to the limits on $\K$.
Since the decay
branching ratios of $H \to W^+W^-/ZZ$ for the SM Higgs 
boson become dominant when $m_H^{~}\gtrsim 200$\,GeV, 
the total width measurement can
impose a strong constraint on $\K$. From the Table\,3 of
Ref.\,\cite{width}, we can derive the accuracy on
the determination of $\kappa$ from the relation
$-n \Delta\Gamma \le \Gamma(\kappa)-\Gamma(\kappa=1)
\le n \Delta\Gamma$,
where $n=1,2$ denote the
$1\sigma$ and $2\sigma$ accuracy, respectively,
$\Gamma(\kappa)$ is the Higgs width for a given
value of $\kappa$, $\Gamma(\kappa=1)$ is the SM Higgs width
and $\Delta\Gamma$ is the expected experimental error of
the width measurement.   We find that at the LHC (with an
integrated luminosity of $300\,\ifb$), measuring the total
Higgs decay width via
\,$p p \to H \to ZZ \to 4\mu $\, can constrain $\K$ as
\beq                     
\ba{lcc}
   {1\sigma\!:}~~~ 0.9 \,\leqq ~|\K|~ \,\leqq 1.1\,,  ~~&~~
   {2\sigma\!:}~~~ 0.8 \,\leqq ~|\K|~ \,\leqq 1.2  \, ,~~~~~
({\rm for}~m_H^{~}=200 \!-\! 300\,{\rm GeV}).
\ea
\label{tevbound2}
\eeq

Before closing this section, a few remarks are in order.
First, when the Higgs boson is lighter than $2m_Z^{~}$,
it is expected that the production
rate of
$qq \to qqH$ with $H \to W W^* \to \ell \ell' {\not \!\!E_T}$
is large enough to be detected at the LHC and
the $HWW$ coupling can be determined
by studying the observables near the Higgs boson resonance~\cite{dieter}.
Similarly, at the future Linear Colliders,
the anomalous $HZZ$ coupling
can be measured via
\,$e^-e^+ \to Z H(\to b\bar{b})$\,\cite{HZZ}.
The implication of these measurements to the determination
of $\K$ will be presented elsewhere\,\cite{ZKHY}.
Finally,
another important effect of an anomalous $\DK$ is
to modify the decay branching ratio of \,$H \to \gamma \gamma$,\,
and consequently, the production rate of
\,$gg \to H \to \gamma \gamma$\, at the LHC.
As a function of $\K$,
Br$[H \to \gamma \gamma]$ decreases when
$\K$ is moving above $1$, and increases otherwise\,\cite{ZKHY}.

\section{\hspace*{-6mm}.$\!$
Conclusions}

In this work, we studied the constraints on the anomalous $HVV$
coupling $\Delta\K$ from the latest precision electroweak data and
from the high energy $V_LV_L$ scatterings at the upcoming LHC
experiments. Comparisons were also made with the other limits
derived from the associate production rate of $VH$ and the total
decay width of the Higgs boson. We showed that the existing
precision data already impose nontrivial  constraints on the
allowed ranges of $\DK$ and the new physics scale $\cut$ (cf.
Fig.\,2 and Table\,1). We further demonstrated that the $V_LV_L$
scatterings, especially $W^+_LW^+_L\to W^+_LW^+_L$ channel, can
sensitively test $\Delta\K$ at the LHC, with an integrated
luminosity of about $300$\,fb$^{-1}$ (cf. Table\,2). Hence,
$V_LV_L$ scatterings are not only important for probing strongly
interacting EWSB mechanism\,\cite{CG,WW95,power} when there is no
light Higgs boson, but also valuable for testing the anomalous
$HVV$ interactions when the Higgs boson is relatively light. It is
particularly important when a positive $\Delta\K$ causes a broad
Higgs resonance that may be hidden by its backgrounds, or when a
negative $\Delta \K$ makes the production rate of a light Higgs
boson too small to be detected near the Higgs resonance. In either
case, because a SM Higgs boson would perfectly saturate the
unitarity, the enhanced $V_LV_L$-scattering signals in the TeV
region directly test the anomalous $HVV$ couplings and thus probe
the underlying EWSB mechanism.

The classic ``no-lose'' theorem
asserts that the LHC, capable of observing the $V_LV_L$
scatterings at $1-2$\,TeV scale, will not fail to test the EWSB
mechanism\,\cite{CG}. Our study provides a new support of this
theorem by showing that even if there exists a Higgs boson as
light as $115-300$\,GeV and some of its on-shell production channels 
may be hard to detect,  the $V_LV_L$ scatterings at the LHC can
still become strong in the TeV regime due to the anomalous $HVV$
interactions. To further discriminate such a non-standard light
Higgs boson from a strongly interacting EWSB sector with no light
resonance will eventually demand a multi-channel analysis at the
LHC by searching for the light resonance through all possible
on-shell production channels, including the gluon-gluon fusion
(unless its interactions with heavy quarks, such as the top and bottom
quarks, are highly suppressed). 
Indeed, the physics with unraveling the
EWSB mechanism could be much more intricate than
naively expected, and studying the $VV$ scatterings at TeV
scale is important for guiding the light Higgs searches via
its on-shell production,  as well as for testing the
EWSB mechanism via $HVV$ interactions
after a light Higgs resonance is found.

\vspace*{3.5mm}
\noindent
{\large {\bf Acknowledgements}}~~~~
We thank
J. Bagger, R. S. Chivukula, P. B. Renton and especially J. Erler
for discussing the precision data,
and D. Zeppenfeld for discussing the weak boson physics of the LHC.
This work was supported by the U.S. DOE grant DE-FG0393ER40757 and
NSF grant PHY-0100677, and by the NSF of China
and Tsinghua Foundation of Fundamental Research.



\begin{thebibliography}{000}

\bibitem{DSB}
For an updated comprehensive review,
C. T. Hill and E. H. Simmons, \tx{hep-ph/0203079}.

\bibitem{susy}
For recent reviews,
H.\,E. Haber, \tx{hep-ph/0212136};
and ``Perspectives on Supersymmetry'',
ed. G.\,L. Kane, World Scientific Publishing Co., 1998.

\bibitem{extraD}
N. Arkani-Hamed, S. Dimopolous, G. R. Dvali,
Phys. Lett. B{\bf 429} (1998) 263;
I. Antoniadis, N. Arkani-Hamed, S. Dimopolous, G. R. Dvali,
Phys. Lett. B{\bf 436} (1998) 257;
L. Randall and R. Sundrum,
Phys. Rev. Lett. {\bf 83} (1999) 3370. 

\bibitem{GK84}
D. B. Kaplan and H. Georgi,
Phys. Lett. B{\bf 136} (1984) 183.

\bibitem{RW}
E.g., T. Rizzo and J. D. Wells,
Phys. Rev. D{\bf 61} (1999) 016007 and references therein.

\bibitem{DeC}
N. Arkani-Hamed, A. G. Cohen, and H. Georgi,
Phys. Lett. B{\bf 513} (2001) 232.

\bibitem{CG}
M. S. Chanowitz and M. K. Gaillard,
Nucl. Phys. B{\bf 261} (1985) 379;
Michael S. Chanowitz, Lecture at Zuoz Summer School
[\tx{hep-ph/9812215}] and references therein.

\bibitem{CCWZ}
C. G. Callen, S. Coleman, J. Wess and B. Zumino,
Phys. Rev. {\bf 177} (1969) 2247.

\bibitem{Wein}
S. Weinberg, Physica A{\bf 96} (1979) 327.

\bibitem{App}
T. Appelquist and C. Bernard,
Phys. Rev. D{\bf 22} (1980) 200;
A. C. Longhitano, Nucl. Phys. B{\bf 188} (1981) 118.

\bibitem{CK}
R. Sekhar Chivukula and V. Koulovassilopoulos,
Phys. Lett. B{\bf 309} (1993) 371.

\bibitem{BW}
W. Buchm\"{u}ller and D. Wyler,
Nucl. Phys. B{\bf 268} (1986) 621.

\bibitem{ZKHY}
B. Zhang, Y.-P. Kuang, H.-J. He, C.-P. Yuan,
in preparation.

\bibitem{NDA}
A. Manohar and H. Georgi,
Nucl. Phys. B{\bf 234} (1984) 189;~
H. Georgi, Phys. Lett. B{\bf 298} (1993) 187
[\tx{hep-ph/9207278}].

\bibitem{PT}
M. E. Peskin and T. Takeuchi,
Phys. Rev. Lett. {\bf 65} (1990) 964;
Phys. Rev. D{\bf 46} (1992) 381.


\bibitem{eg}
E.g., S. Dawson and G. Valencia,
Nucl. Phys. B{\bf 439} (1995) 3.

\bibitem{BL}
H. Georgi,
Ann. Rev. Nucl. \& Part. Sci. {\bf 43} (1994) 209,
and references therein.


\bibitem{PDG}
Particle Data Group (K. Hagiwara {\it et al}.), Phys. Rev. D {\bf 66}
(2002) 010001.

\bibitem{seesaw}
B.\,A. Dobrescu, C.\,T. Hill, Phys. Rev. Lett. {\bf 81} (1998) 2634;
R.\,S. Chivukula, B.\,A. Dobrescu, H. Georgi, C.\,T. Hill,
Phys. Rev. D{\bf 59} (1999) 075003;
H.--J. He, C. T. Hill, T. Tait, Phys. Rev. D{\bf 65} (2002) 055006;
H.--J. He, T. Tait, C.--P. Yuan, 
Phys. Rev. D{\bf 62} (2000) 011702 (R).

\bibitem{bagger}
J. Bagger, A. F. Falk and M. Swartz,
Phys. Rev. Lett. {\bf 84} (2000) 1385;
R. S. Chivukula, C. Holbling and N. Evans,
Phys. Rev. Lett. {\bf 85} (2000) 511;
C. Kolda and H. Murayama, JHEP {\bf 0007} (2000) 035;
M. E. Peskin and J. D. Wells, Phys. Rev. D{\bf 64} (2001) 093003.

\bibitem{HPS}
H.--J. He, N. Polonsky and S. Su, Phys. Rev. D{\bf 64} (2001) 053004
[\tx{hep-ph/0102144}].

\bibitem{snow}
R. Sekhar Chivukula and C. Holbling,
Snowmass contribution [\tx{hep-ph/0110214}].

\bibitem{Mike}
M. S. Chanowitz,
Phys. Rev. D{\bf 66} (2002) 073002 [\tx{hep-ph/0207123}].

\bibitem{NuTeVx}
E.g., G. A. Miller and A. W. Thomas,
\tx{hep-ex/0204007}; S. Davidson, \tx{hep-ph/0209316}.

\bibitem{elgapp}
J. Erler, private communications, \tx{hep-ph/0005084}
and review at http://pdg.lbl.gov.

\bibitem{data-new}
M. W. Gr\"{u}newald, talk at ICHEP, Amsterdam,
July 24-31, 2002 [\tx{hep-ex/0210003}];
see also, P. B. Renton, Rept. Prog. Phys.
{\bf 65} (2002) 1271 [\tx{hep-ph/0206231}].


\bibitem{power}
H.--J. He, Y.--P. Kuang and C.--P. Yuan,
Phys. Rev. D{\bf 55} (1997) 3038 [\tx{hep-ph/9611316}];
Phys. Lett. B{\bf 382} (1996) 149 [\tx{hep-ph/9604309}];
and review in DESY-97-056 [\tx{hep-ph/9704276}].

\bibitem{nsH2}
R. S. Chivukula, M. J. Dugan, M. Golden,
Phys. Lett. B{\bf 336} (1994) 62
[\tx{hep-ph/9406281}].

\bibitem{cteq6}
J. Pumplin, {\it et al.,}
JHEP {\bf 0207} (2002) 012
[\tx{hep-ph/0201195}].

\bibitem{WW95}
J. Bagger, V. Barger, K. Cheung, J. Gunion, T. Han, G. Ladinsky,
R. Rosenfeld, C.--P. Yuan,
Phys. Rev. D{\bf 52} (1995) 3878 [\tx{hep-ph/9504426}].

\bibitem{Yuan}
C.--P. Yuan,
{\it Proposals for Studying TeV $W_LW_L\to W_LW_L$ Interactions
Experimentally}, [\tx{hep-ph/9712513}],
in {\it Perspectives on Higgs Physics,}
G. L. Kane (Ed.), World Scientific Pub.

\bibitem{agrawal}
P. Agrawal, Mod. Phys. Lett. A{\bf 16} (2001) 897
[\tx{hep-ph/0011347}].

\bibitem{width}
V. Drollinger and A. Sopczak,
in the proceedings of LCWS-2001 [\tx{hep-ph/0102342}].

\bibitem{dieter}
D. Zeppenfeld, Snowmass contribution [\tx{hep-ph/0203123}]
and references therein.

\bibitem{HZZ}
E.g.,
K. Hagiwara, {\it et al.,}
Eur. Phys. J. {\bf 14} (2000) 457,
and references therein.



\end{thebibliography}
\end{document}